\documentclass{emulateapj}
\slugcomment{ApJL, in prep.}
\shorttitle{A new NGC~253 dwarf galaxy}
\shortauthors{Sand et al.}

\begin{document}
 \title{Discovery of a new faint dwarf galaxy associated with NGC~253~$\!$\altaffilmark{*}}

\author{D. J. Sand,$\!$\altaffilmark{1} D. Crnojevi\'{c},$\!$\altaffilmark{1} J. Strader,$\!$\altaffilmark{2} E. Toloba,$\!$\altaffilmark{3,4} J.D. Simon,$\!$\altaffilmark{4} N. Caldwell,$\!$\altaffilmark{5} P. Guhathakurta,$\!$\altaffilmark{3} B. McLeod,$\!$\altaffilmark{5} A. C. Seth~$\!$\altaffilmark{6}} \email{david.sand@ttu.edu}

\begin{abstract}
We report the discovery of a new faint dwarf galaxy, which we dub Scl-MM-Dw1, at a projected distance of $\sim$65 kpc from the spiral galaxy NGC~253. The discovery results from the Panoramic Imaging Survey of Centaurus and Sculptor (PISCeS), a program with the Magellan/Megacam imager to study faint substructure in resolved stellar light around massive galaxies outside of the Local Group.  We measure a tip of the red giant branch distance to Scl-MM-Dw1 of $D$=3.9$\pm$0.5 Mpc, consistent with that of NGC~253, making their association likely.  The new dwarf's stellar population is complex, with an old, metal poor red giant branch ($\gtrsim$10 Gyr, [Fe/H]$\sim$$-$2), and an asymptotic giant branch with an age of $\sim$500 Myr.  Scl-MM-Dw1 has a half-light radius of $r_{h}$=340$\pm$50 pc and an absolute magnitude of $M_{V}$=$-$10.3$\pm$0.6 mag, comparable to the Milky Way's satellites at the same luminosity.  Once complete, our imaging survey of NGC~253 and other nearby massive galaxies will provide a census of faint substructure in halos beyond the Local Group, both to put our own environment into context and to confront models of hierarchical structure formation.

\end{abstract}
\keywords{later}
 
\altaffiltext{*}{This paper includes data gathered with the 6.5 meter Magellan Telescopes located at Las Campanas Observatory, Chile.}
\altaffiltext{1}{Texas Tech University, Physics Department, Box 41051, Lubbock, TX 79409-1051, USA}
\altaffiltext{2}{Michigan State University, Department of Physics and Astronomy, East Lansing, MI 48824, USA}
\altaffiltext{3}{UCO/Lick Observatory, University of California, Santa Cruz, 1156 High Street, Santa Cruz, CA 95064, USA}
\altaffiltext{4}{Observatories of the Carnegie Institution for Science, 813 Santa Barbara Street, Pasadena, CA 91101, USA}
\altaffiltext{5}{Harvard-Smithsonian Center for Astrophysics, Cambridge, MA 02138, USA}
\altaffiltext{6}{Department of Physics and Astronomy, University of Utah, Salt Lake City, UT 84112, USA}

\section{Introduction}


The $\Lambda$+Cold Dark Matter ($\Lambda$CDM) model for structure formation is successful on large scales \citep[$\gtrsim$10 Mpc; e.g.][]{Efstathiou92,Jaffe01,Percival01,Spergel07}. In this canonical model, galaxies grow hierarchically within dark matter halos \citep[e.g.][]{Springel06}. However, a quantitative verification of this model has met with extensive challenges, particularly on galaxy scales and smaller. For instance, comparisons between the number of subhalos seen in numerical simulations and the number of Milky Way or M31 dwarf galaxies show a factor of $>$100 discrepancy \citep[the ``missing satellites problem", e.g.][]{Klypin99,M99b}. Further, kinematic studies of the most luminous Local Group dwarfs indicate that their densities are lower than those of simulated halos of the same mass  \citep[the ``too big to fail" problem;][]{Boylan11,Boylan12}.
~~These small-scale tests of $\Lambda$CDM galaxy formation models have been mostly confined to substructure around the Milky Way and M31.

Two spiral galaxy halos in a loose group do not sample the full spectrum of galaxy masses, morphologies and environments necessary to have a complete picture of galaxy assembly. In particular, we need to
determine whether the Local Group just happens to be an outlier in terms of its faint dwarf galaxy population, or if substructure in other galaxies presents similar tension with $\Lambda$CDM.  
Pioneering studies of other galaxies have begun, including resolved stellar population work in M81 \citep{Chiboucas09,Chiboucas13} and low surface brightness searches out to larger distances \citep[e.g.][]{Merritt14}. 

In this work, we present the discovery of the faint dwarf galaxy Scl-MM-Dw1, the first result of a new survey to study faint substructure around NGC~253 (a star-bursting spiral galaxy in the Sculptor group) and NGC~5128 (the dominant elliptical galaxy in the Centaurus A group) in order to broaden our observational knowledge of faint substructure beyond the Local Group.  In \S~\ref{sec:survey_desc}, we briefly summarize the strategy for our panoramic imaging survey of nearby galaxies.  We present our data reduction methodology in \S~\ref{sec:datareduce} and the physical properties of Scl-MM-Dw1 in \S~\ref{sec:properties}.  We discuss and conclude in \S~\ref{sec:conclude}.

\section{Survey Description \& Discovery of Scl-MM-Dw1}\label{sec:survey_desc}

We have begun the Panoramic Imaging Survey of Centaurus and Sculptor (PISCeS), a resolved stellar survey of two of our nearest massive galaxy neighbors ---NGC 253 and NGC 5128---to search for archaeological remnants of their buildup via tidal streams and satellite dwarf galaxies down to $M_V\sim-8$ mag.  The most recent surveys in these galaxies for dwarf galaxy companions relied on photographic plates, and suffered from incompleteness fainter than $M_{V}$$\sim$$-$14 mag \citep{Cote97,Jerjen00,K02_cen,K04}. Our goal is to survey each galaxy out to $r$=150 kpc to allow a direct comparison with the PAndAS survey in M31 \citep{McConnachie09}, and recent work around the Milky Way.

To accomplish this ambitious task, we are utilizing the Megacam instrument \citep{McLeod06}, which has a total field of view of $\sim$24'$\times$24' at the $f$/5 focus on the Magellan Clay telescope. The outer survey radius of $\sim$150 kpc amounts to $\sim$21 square degrees on the sky for NGC~253/NGC~5128 (both at $D$$\sim$3.5 Mpc), for a total of $\sim$130 Megacam pointings for each halo. In rough accordance with previous dwarf galaxy discoveries in the Sculptor group, we will label our new dwarfs numerically starting with Scl-MM-Dw1, where the ``MM" denotes the Magellan/Megacam instrumentation utilized by our program.

Imaging is done in two bands, $g$ and $r$, chosen for survey speed and so that new dwarf candidates can be clearly identified and initial properties (e.g. distance, size, approximate star formation history, and luminosity) ascertained from their resolved stellar populations during the course of the survey. At the distance to NGC~253/NGC~5128, the tip of the red giant branch (TRGB) is at $r$$\sim$24.5 mag; our simulations suggested that reaching $M_{V}$$\sim$$-$8 dwarf galaxies requires imaging depths of $r$,$g$$\sim$26.5 mag (to be presented in a forthcoming work).  In good seeing conditions ($\lesssim$0\farcs8), this requires 30 minutes of integration in each band.



\begin{figure*}
\begin{center}
\mbox{ \epsfysize=7.0cm \epsfbox{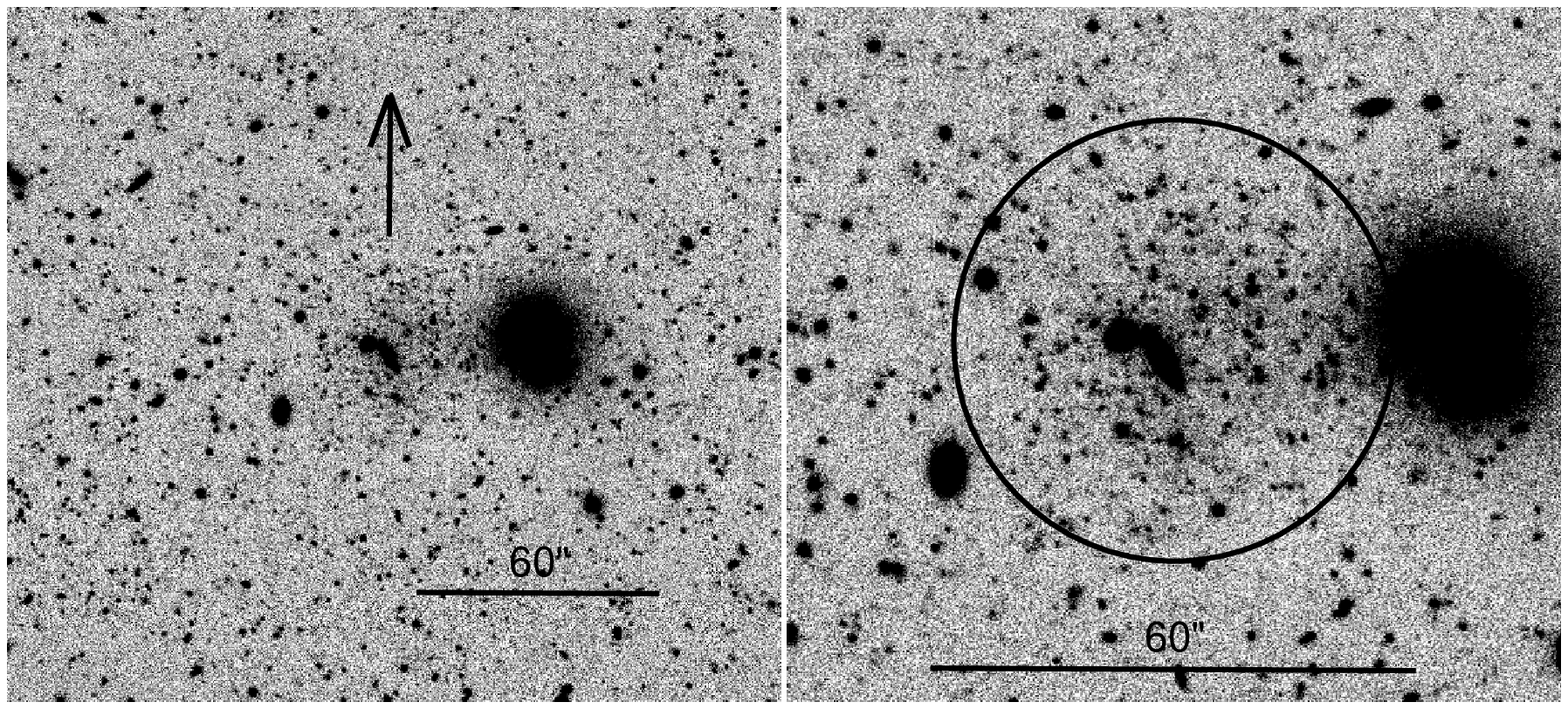}} 
\mbox{ \epsfysize=10.0cm \epsfbox{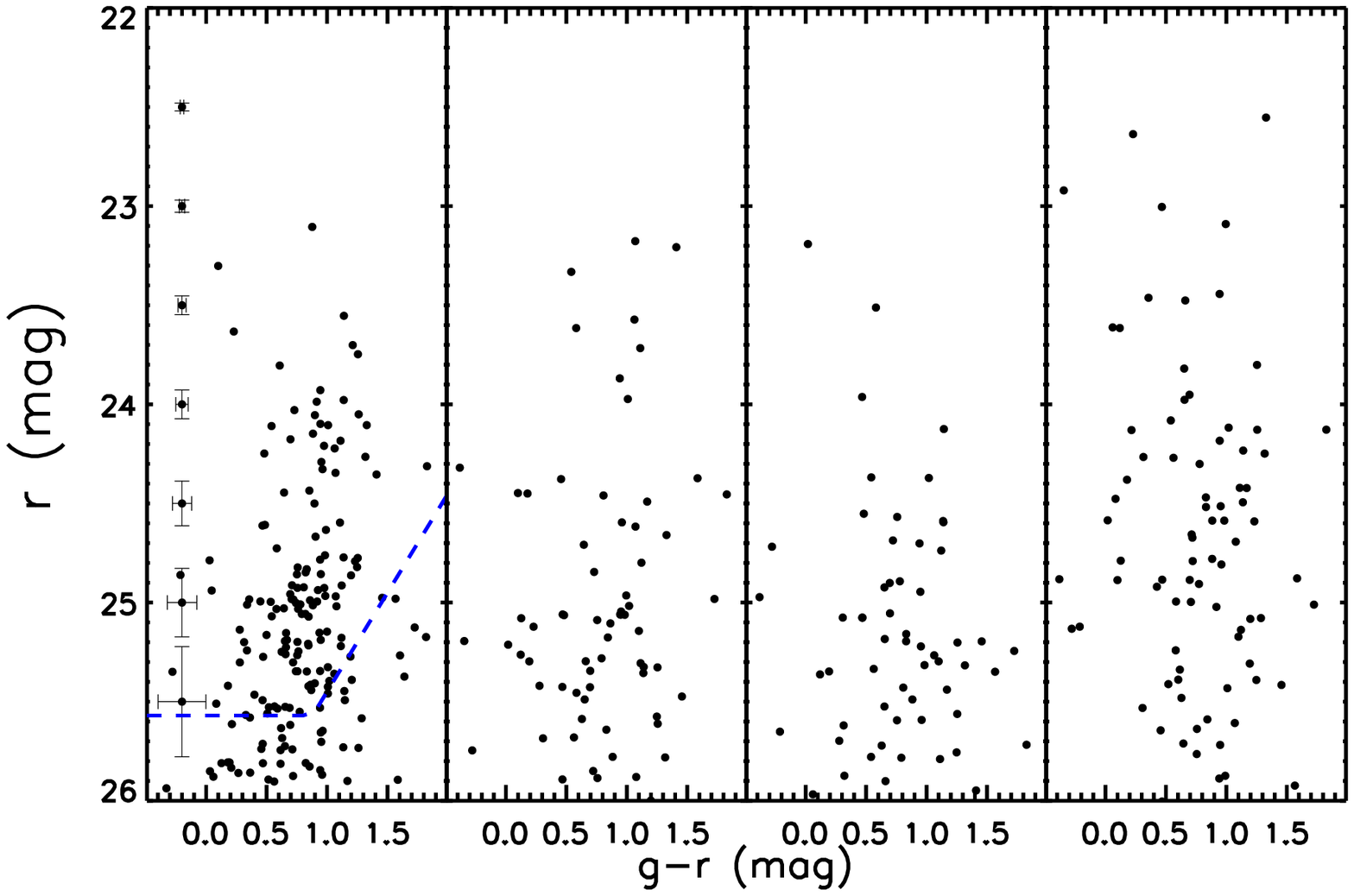}} 
\caption{  {\bf Top:} Scl-MM-Dw1 as seen in our $g$-band image.  The top left panel shows a zoomed out view, with the black arrow indicating the direction towards NGC~253 (nearly due North) and $\sim$65 kpc away in projection.  There is also a clear surface brightness enhancement associated with the dwarf. The top right panel shows a zoomed in view, with many resolved stars apparent.  Note also the spiral/S0 galaxy and a bright foreground star near the center of Scl-MM-Dw1. The radius of the circle is 0.45 arcmin, which is the region from which the CMD in the bottom panel is drawn.  {\bf Bottom:} The bottom left panel shows the CMD of Scl-MM-Dw1 within the 0.45 arcminute radius circle drawn.  There is a clear overdensity of stars in this region, with a morphology similar to a RGB, which we analyze in more detail in \S~\ref{sec:stellarpop}.  Along the left side of the CMD are the typical uncertainties at different $r$-band magnitudes, as determined via our artificial star tests.  The blue dashed line shows the 50\% completeness limit.  The three adjacent panels show CMDs from random equal-area regions of the same Megacam field, illustrating typical ``background" CMDs.  \label{fig:dwarf}}
\end{center}
\end{figure*}

During the Fall 2012 observing season, we identified a clear dwarf galaxy candidate around NGC~253 using data taken with poor seeing. New, deeper observations were taken in Fall 2013, in much better seeing conditions  (0\farcs5--0\farcs6; Figure~\ref{fig:dwarf}).  This new dwarf galaxy, which we dub  Scl-MM-Dw1, is $\sim$65 kpc away from the center of NGC~253 in projection.  Scl-MM-Dw1 is just visible in Digital Sky Survey archival images, but at a level where a detection is not secure, especially due to the presence of a handful of bright foreground stars and background galaxies in its vicinity. 

\section{Data Reduction} \label{sec:datareduce}

The data presented in this paper were taken on 2013 September 2 and 7 (UT) with Magellan/Megacam. Photometric conditions were variable, but the seeing was excellent.  The final stacked images consisted of 15$\times$300s images in both the $g$ and $r$ band, with image point spread functions (PSFs) of 0\farcs5--0\farcs6.  Initial data reduction, including standard image detrending, astrometric matching and stacking is performed by the Smithsonian Astrophysical Observatory Telescope Data Center utilizing a code developed by M. Conroy, J. Roll and B. McLeod.  Stellar photometry on the final image stacks was performed using a methodology similar to that of previous work utilizing Magellan/Megacam for resolved stellar studies \citep{Sand12} with the {\sc DAOPHOTII/Allstar} package \citep{Stetson94}.


Our Fall 2013 data of Scl-MM-Dw1 were not taken in photometric conditions, and so we utilized data taken in Fall 2012 of the same field in clear conditions (but poor seeing) to calibrate the presented data.  We convert instrumental magnitudes into the SDSS photometric system by observing fields covered by SDSS at different airmasses during clear nights, which allowed for an atmospheric extinction term to be measured.  Zeropoints and color terms were similar to our previous work with Magellan/Megacam \citep{Sand12}.  

Once the Fall 2013 data was calibrated, we performed a series of artificial star tests (utilizing the {\sc DAOPHOT} routine {\sc ADDSTAR}) to calculate our photometric errors and completeness as a function of magnitude and color for the field.
Since crowding is not an issue, artificial stars were placed into our images on a regular grid whose spacing was several times that of the stellar PSF.  Over ten iterations, we injected $\sim$6$\times$10$^5$ artificial stars into our data, covering a $r$-band magnitude range of 18 to 30 and a $g-r$ color of $-$0.5 to 2.0.  Using the exact same photometric pipeline as was used on the unaltered data, we find a 50\% (90\%) percent completeness level at $r$=25.6 (24.4) and $g$=26.5 (25.3). These limits are slightly shallower than our ultimate goals due to the sky conditions; data taken on clear, good-seeing nights do hit our magnitude targets as described in \S~\ref{sec:survey_desc}.  There is no indication that the unresolved light from Scl-MM-Dw1 changes our completeness or uncertainties in that region.  

Our final catalog was corrected for Galactic extinction \citep{Schlafly11}, and all magnitudes reported in the remainder of this paper will be corrected in this way.  Figure~\ref{fig:dwarf} shows our color magnitude diagram (CMD) of Scl-MM-Dw1, where the error bars show the typical magnitude and color uncertainty at different $r$-band magnitude levels, and a $g-r$ of 0.8 mag.

\section{Properties of Scl-MM-Dw1}\label{sec:properties}

In this section, we determine the basic properties of Scl-MM-Dw1.  We lead the discussion with its stellar populations, as these are critical for the subsections that follow: the distance, structure, and luminosity of the newly discovered dwarf.
\subsection{Stellar Population} \label{sec:stellarpop}
Close inspection of the CMD as seen in Figure~\ref{fig:dwarf} hints at two interpretations for the stellar population and approximate distance to Scl-MM-Dw1, and we critically assess each here.  

\begin{figure*}
\begin{center}
\mbox{ \epsfysize=10.0cm \epsfbox{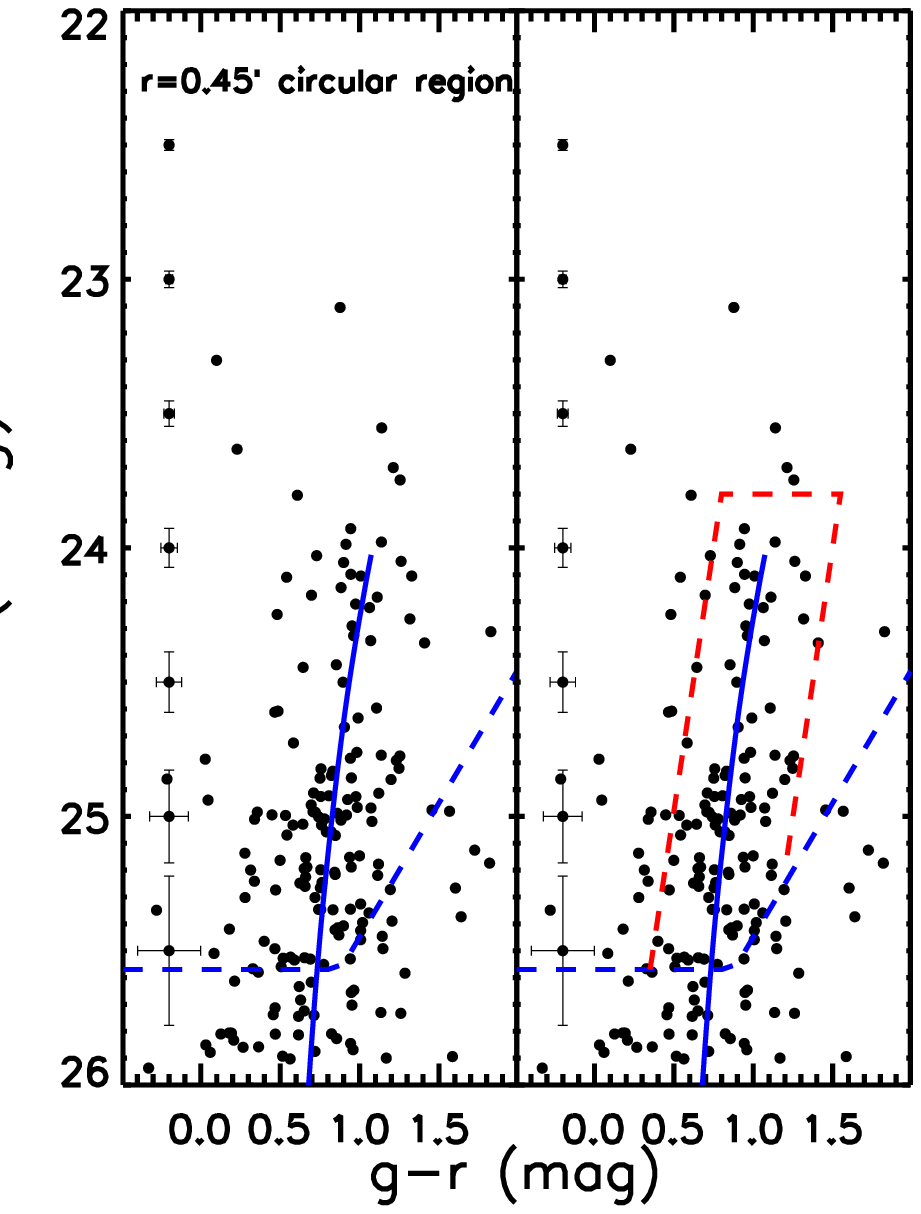}} 
\mbox{ \epsfysize=7.0cm \epsfbox{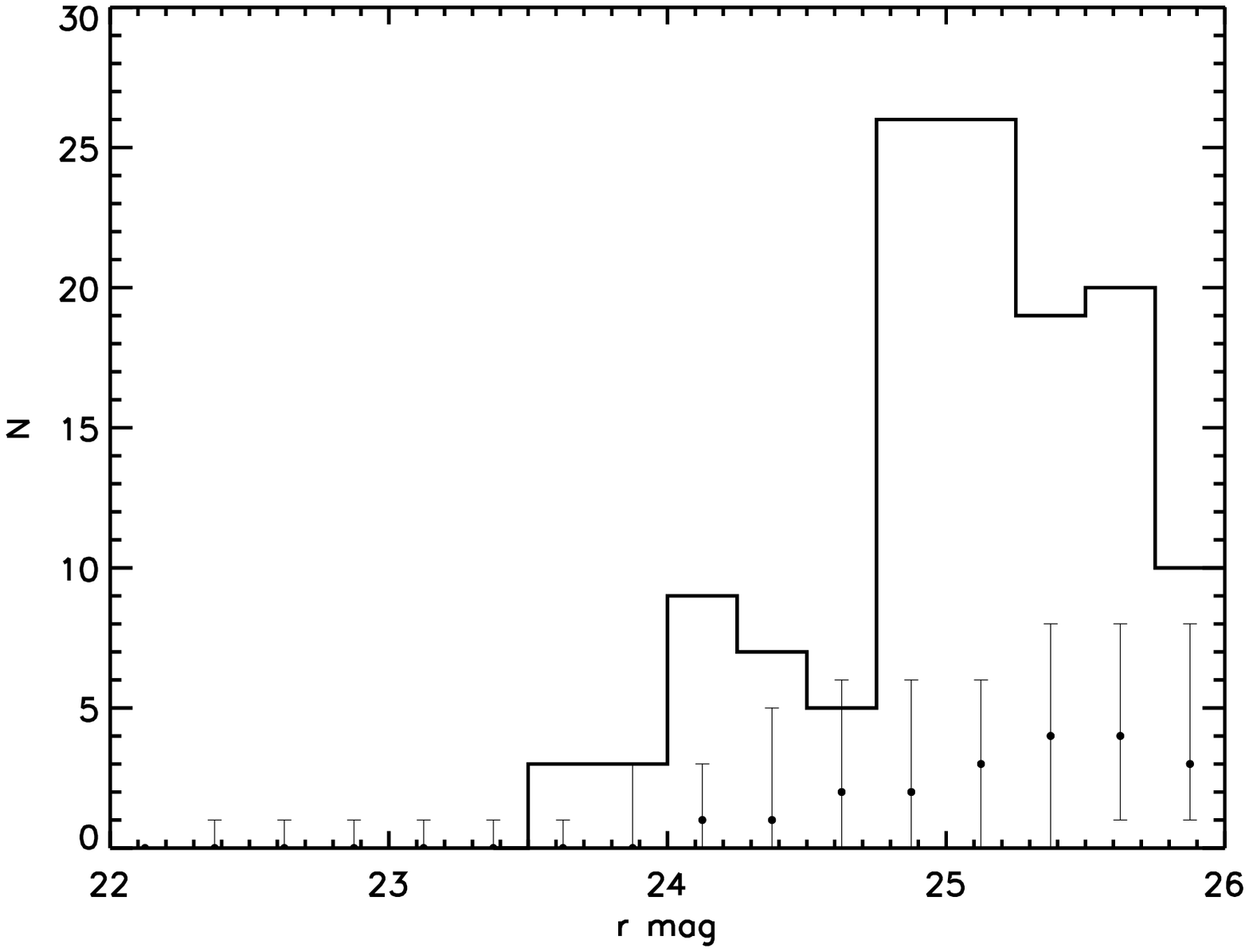}} 
\caption{ The ``nearby" interpretation of Scl-MM-Dw1, which we ultimately reject.  {\bf Left:} A CMD of Scl-MM-Dw1, identical to that shown in Fig~\ref{fig:dwarf}, is shown.  The blue theoretical isochrone is of a 10 Gyr, [Fe/H]=$-$2.0 stellar population \citep{Dotter08} at a distance of $\mu$=27.1 ($D$=2.6 Mpc), $\Delta \mu$$\sim$0.6 mag closer than NGC~253.  There is a gap in star counts at $r$$\approx$24.5 mag and a $g-r$$\approx$0.8 mag, which would be unusual for a RGB.  We plot the luminosity function of this scenario's RGB in the right panel, based on stars in the red dashed region.  {\bf Right:} Luminosity function of stars consistent with being a part of the RGB of Scl-MM-Dw1, if it was at a distance of $\mu$=27.1.   The dotted points show the luminosity function of equal-area random pointings throughout the field with the error bars spanning 95\% of the random draws (see \S~\ref{sec:stellarpop} for details).  The secondary peak in Scl-MM-Dw1's luminosity function at $r$$\sim$24 is statistically significant; the stars associated with this secondary peak are also clustered around the position of Scl-MM-Dw1, see Figure~\ref{fig:cmd_right}.  The number counts of stars in the gap between the two peaks in the luminosity function are consistent with being background.  Based on this, we are confident that Scl-MM-Dw1 is not a dwarf galaxy at $\mu$=27.1 ($D$=2.6 Mpc), but that the secondary peak is of an AGB stellar population.  \label{fig:cmd_close}}
\end{center}
\end{figure*}

The first scenario---which we ultimately argue against---places Scl-MM-Dw1 nearer than the nominal distance to NGC~253 (with a difference in distance modulus of $\Delta \mu$$\sim$0.6 mag) at $\mu$$\sim$27.1 mag ($D$=2.6 Mpc), as shown in Figure~\ref{fig:cmd_close}.  In the left panel of Figure~\ref{fig:cmd_close} we over-plot a 10 Gyr, [Fe/H]=$-$2.0 theoretical isochrone \citep{Dotter08} at this distance, showing a reasonable match to the overdensity of stars.  However, if true, the morphology of the RGB would be odd, as there appears to be a gap in the stellar luminosity function at $r$$\approx$24.5 mag and a color of $g-r$$\approx$0.8 mag.  In the right panel of Figure~\ref{fig:cmd_close} we plot a luminosity function of stars consistent with the RGB of this ``nearby" scenario, along with a sample of luminosity functions of random, equal-area fields in the same Megacam pointing.~~
It is clear that the number of stars in the gap is consistent with being a foreground population, but the secondary peak in the luminosity function at $r$$\sim$24 mag is statistically significant.  An RGB luminosity function would not have such a gap, which suggests that the true dwarf distance is further away, and that the secondary peak is not the TRGB, but a different stellar population associated with the dwarf. 

\begin{figure*}
\begin{center}
\mbox{ \epsfysize=10.cm \epsfbox{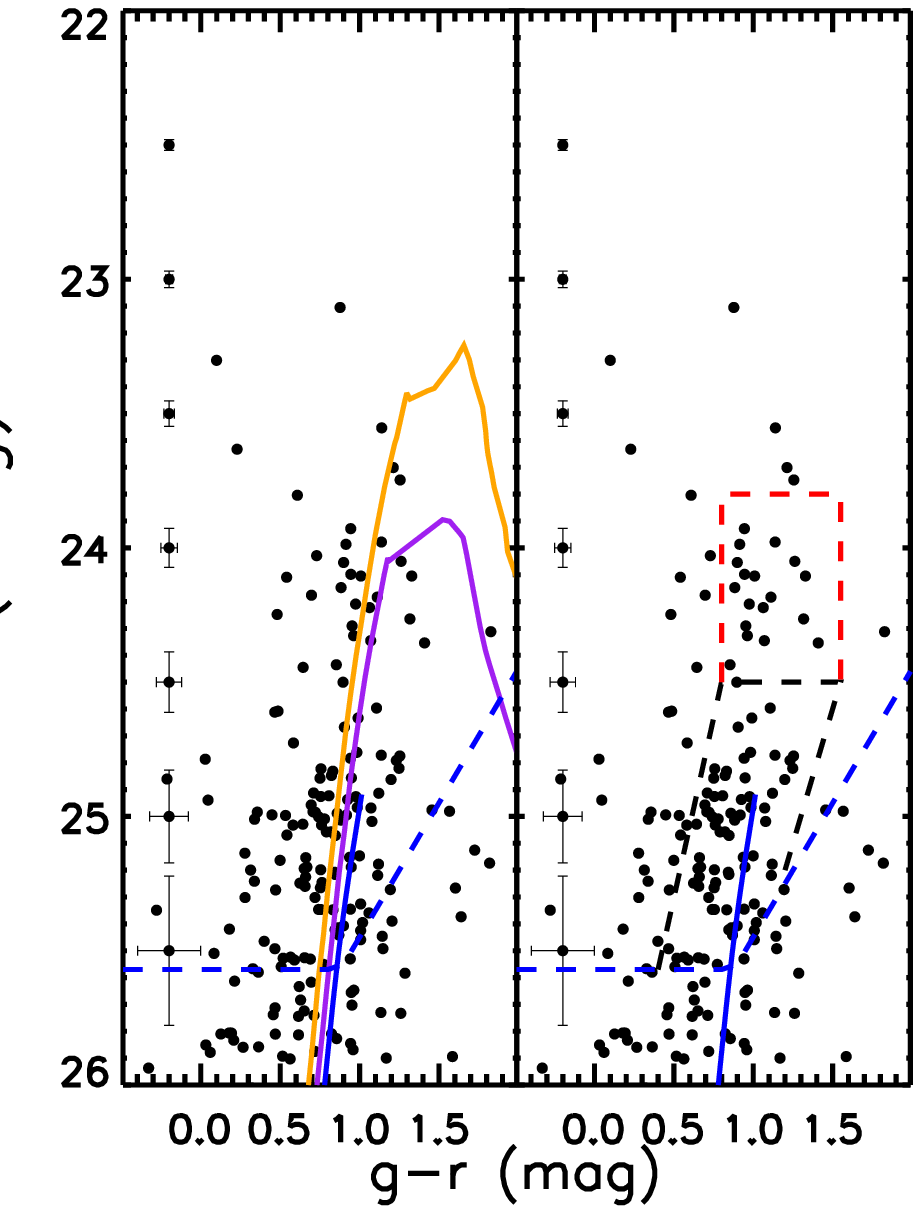}} 
\mbox{ \epsfysize=8.0cm \epsfbox{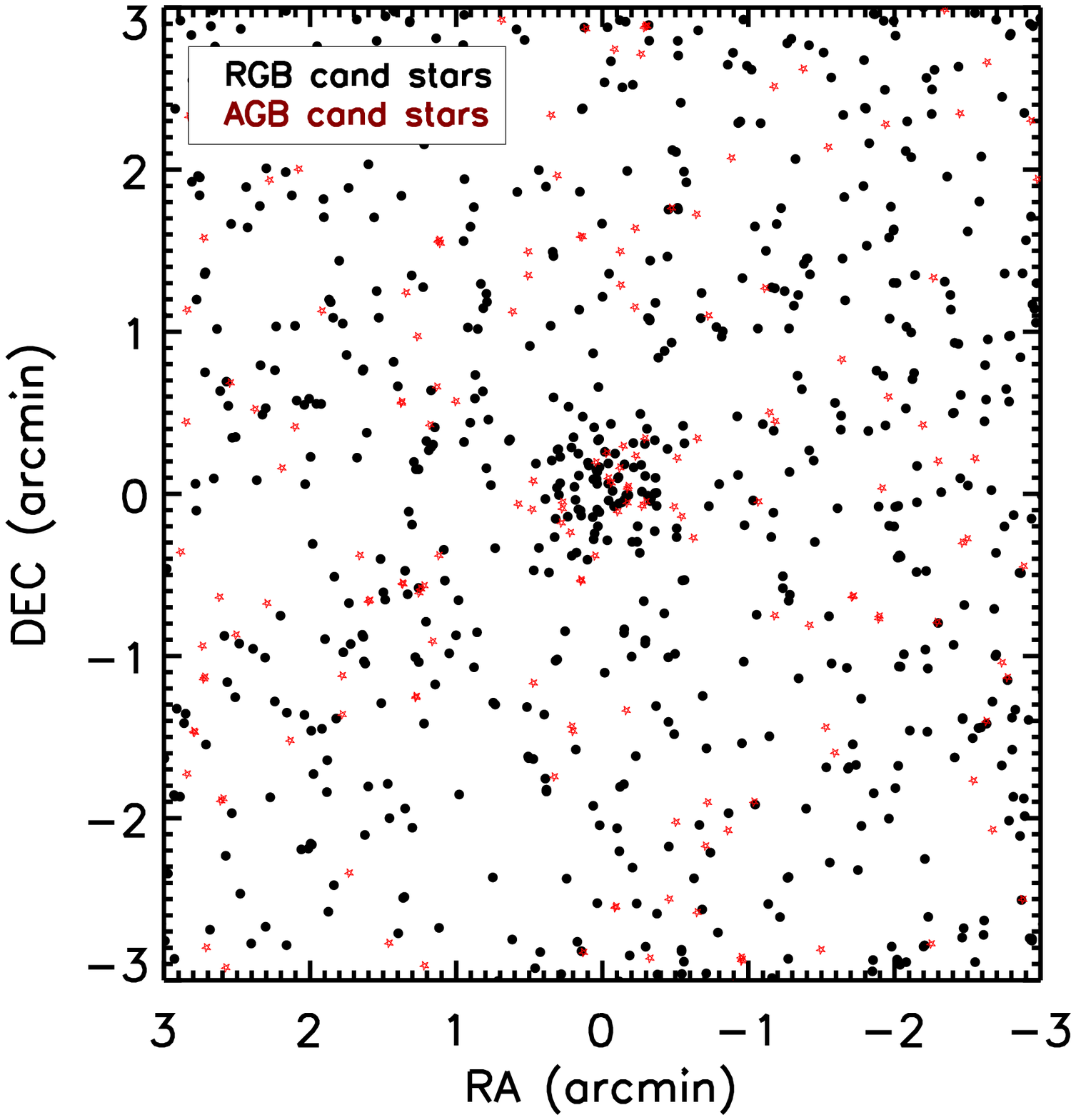}} 
\caption{The NGC~253 satellite interpretation of Scl-MM-Dw1.  {\bf Left:} The CMD as plotted in Figures~\ref{fig:dwarf} \& \ref{fig:cmd_close}. The orange and purple isochrones are for a 400 and 630 Myr stellar population with a [Fe/H]=$-$1.  The blue isochrone is of a 13 Gyr, [Fe/H]=$-$2.0 stellar population. {\bf Right:}  Spatial map of stars consistent with being AGB (red; red selection box in the left panel) and RGB (black; black selection box in the left panel) stars. There is an overdensity of stars of both populations at the position of the dwarf. \label{fig:cmd_right}}
\end{center}
\end{figure*}

This leads to the second scenario, where Scl-MM-Dw1 is consistent with being at the distance of NGC~253 (we directly measure the distance to Scl-MM-Dw1 via its TRGB in \S~\ref{sec:distance}).  This is illustrated in Figure~\ref{fig:cmd_right}, where we over-plot a 13 Gyr, [Fe/H]=$-2.0$ theoretical isochrone \citep[in blue;][]{Dotter08}, along with two younger isochrones of 400 and 630 Myr (purple and orange, respectively) and [Fe/H]=$-$1.0 \citep[here we use the isochrones of][as special attention was paid to the AGB stellar population]{Girardi10}.  In this case, the old RGB is consistent with the main peak in the luminosity function, at roughly the distance of NGC~253, while the secondary peak could be from a younger, AGB stellar population of $\sim$500 Myr.  

We are confident that at least some of the stars in this secondary peak in the CMD/luminosity function are associated with Scl-MM-Dw1 because they appear to be spatially clustered as the RGB stars are.  In the right panel of Figure~\ref{fig:cmd_right}, we show a spatial map of stars consistent with the RGB and AGB stellar populations in this scenario.  Both populations show a clear overdensity at the position of the dwarf, reaffirming their significance.  

Only a broad view of the stellar population in Scl-MM-Dw1 can be gleaned from the current data set.  Overall, the CMD of Scl-MM-Dw1 is consistent with an old ($\gtrsim$10 Gyr), metal poor ([Fe/H]$\sim$$-$2) stellar population at roughly the distance to NGC~253 (see \S~\ref{sec:distance}), with evidence for a younger ($\sim$500 Myr) stellar population as well.  A low level of even more recent star formation could also be present, but would not be detected given our current image depths.  A similar CMD morphology (with an RGB, `gap' and second brighter stellar population) is often seen in the literature for other dwarfs.  Indeed, inspection of the CMD compilation of \citet{Dalcanton09} shows several dwarfs with a similar morphology (including DDO44/KK61, KDG2/E540-30, and KDG61/KK81) where an old RGB stellar population exists with a younger AGB stellar population.  Deeper optical and near-infrared imaging will be able to further constrain Scl-MM-Dw1's stellar population.




\subsection{The Distance to Scl-MM-Dw1}\label{sec:distance}

We measure the distance of Scl-MM-Dw1 with the TRGB method
\citep[e.g.,][]{lee93,rizzi07}. For metal-poor populations, the TRGB has a constant
$I$-band absolute magnitude, making it a widely used distance indicator. 
However, to our best knowledge there is no empirical TRGB calibration for 
SDSS bands \citep[e.g.,][]{bellazzini08}. We thus adopt the Dartmouth stellar
evolutionary models \citep{Dotter08} and derive the theoretical absolute TRGB 
magnitude in $r$-band using isochrones with a fixed age of 12~Gyr and metallicities 
ranging from [Fe/H]$=-2.5$ to $-1.0$, obtaining $M_r^{TRGB}=-3.01\pm0.1$.
We apply a Sobel edge-detection filter \citep{lee93} to the luminosity function
of Scl-MM-Dw1 with a color cut of $0.5<(g-r)<1.3$, in order to minimize contamination
from background galaxies and foreground stars. The resulting value is 
$r_{TRGB}=24.97\pm0.26$, which translates into a distance modulus
of $m-M$=$27.98\pm0.32$ with our TRGB calibration. The uncertainties reflect the 
photometric errors and the sparseness of the luminosity function.
As a test, we also compute $r_{TRGB}$ for all stars in the Megacam
pointing not belonging to Scl-MM-Dw1, i.e., probable NGC~253 halo stars.
We find $24.69\pm0.24$ which gives a distance modulus of $m-M$=$27.70\pm0.30$. The excellent 
agreement with literature estimates for the distance of NGC~253 
\citep[e.g.,][who found $m-M$=27.70$\pm$0.07]{ghosts} confirms the robustness of our adopted calibration. 
We conclude that Scl-MM-Dw1 lies at approximately the same distance as NGC~253 
within the uncertainties. 

\subsection{Structure}
Despite its distance and relatively small number of resolved stars, we can determine many of the structural parameters of Scl-MM-Dw1 through standard methods.  We use a maximum likelihood technique for constraining structural parameters, based on the recipe of \citet{sdssstruct}, and identical to that done in \citet{Sand12}. We cut out a 10$\times$10 arcmin$^{2}$ region around the approximate center of the dwarf for use in our calculations. The stars selected for the structural analysis are those in the boxed regions of Figure~\ref{fig:cmd_right} (left panel), corresponding to the RGB and AGB star candidates in Scl-MM-Dw1.  We fit a standard exponential profile plus constant background to the data, with the following free parameters: the central position ($\alpha_{0}$,$\delta_{0}$), position angle (PA; $\theta$), ellipticity ($\epsilon$), half-light radius ($r_{h}$) and background surface density ($\Sigma_{b}$). Uncertainties on structural parameters were determined via bootstrap resamples, from which 68\% confidence limits were calculated.  

Our results are presented in Table~\ref{table:properties}.  The half light radius of Scl-MM-Dw1 is 340$\pm$50 pc, which is comparable to the size of the MW's classical dwarf spheroidals.  The central position of Scl-MM-Dw1 is also well constrained.  The ellipticity (and thus position angle) of Scl-MM-Dw1 is not well constrained, and our maximum likelihood analysis can provide only an upper limit of $\epsilon$$<$0.42 (95\% confidence limit), due to the sparse number of stars inferred to be Scl-MM-Dw1 members ($\sim$100) in the current data set.

\subsection{Luminosity}

We estimate Scl-MM-Dw1's luminosity directly from aperture photometry.  We use an aperture radius equal to the half-light radius (16\farcs8), and randomly place 100 equal-area apertures throughout our Megacam field of view to estimate the background.  
This method has the advantage that it should both account for the clear background galaxies (and foreground stars) in our Scl-MM-Dw1 aperture in a statistical sense, as well as any halo NGC~253 stars.  Since by definition an aperture equal to the half-light radius only includes half of the flux from the dwarf, we correct the dwarf flux by a factor of two. Using this methodology we find $M_{g}$=$-$10.0$\pm$0.5 and $M_{r}$=$-$10.5$\pm$0.6, where the uncertainty on the luminosity was found based on the scatter in measurements from the 100 background apertures, as well as the uncertainty in our measured distance modulus (see \S~\ref{sec:distance}).  To convert from $M_{r}$, $M_{g}$ magnitudes to $M_{V}$, we use the filter transformation equations of \citet{Jordi06}, and find $M_{V}$=$-$10.3$\pm$0.6.

\section{Discussion and Conclusions}\label{sec:conclude}

We have presented the discovery of Scl-MM-Dw1, a faint dwarf galaxy found $\sim$65 kpc in projection from NGC~253. Its inferred distance (3.9$\pm$0.5 Mpc) matches NGC~253 itself, making a physical association likely.
The stellar population of Scl-MM-Dw1 is complex, with an RGB consistent with an old, metal poor stellar population ($>$10 Gyr; [Fe/H]$\sim$$-$2) and AGB stars with an age of $\sim$500 Myr. Dwarfs with recent star formation are rarely seen within $\sim$250 kpc of massive parent galaxies such as NGC~253 \citep[e.g.][among many others]{Grebel03,Grcevich09}, so an improved distance to the dwarf would be of significant interest.
Scl-MM-Dw1 is not detected in the HI Parkes All-Sky Survey \citep[HIPASS;][]{Barnes01}, with a 3-$\sigma$ HI gas mass upper limit of $M_{HI}$$\lesssim$1$\times$10$^{7}$ $M_{\odot}$.

\begin{figure*}
\begin{center}
\mbox{ \epsfysize=10.cm \epsfbox{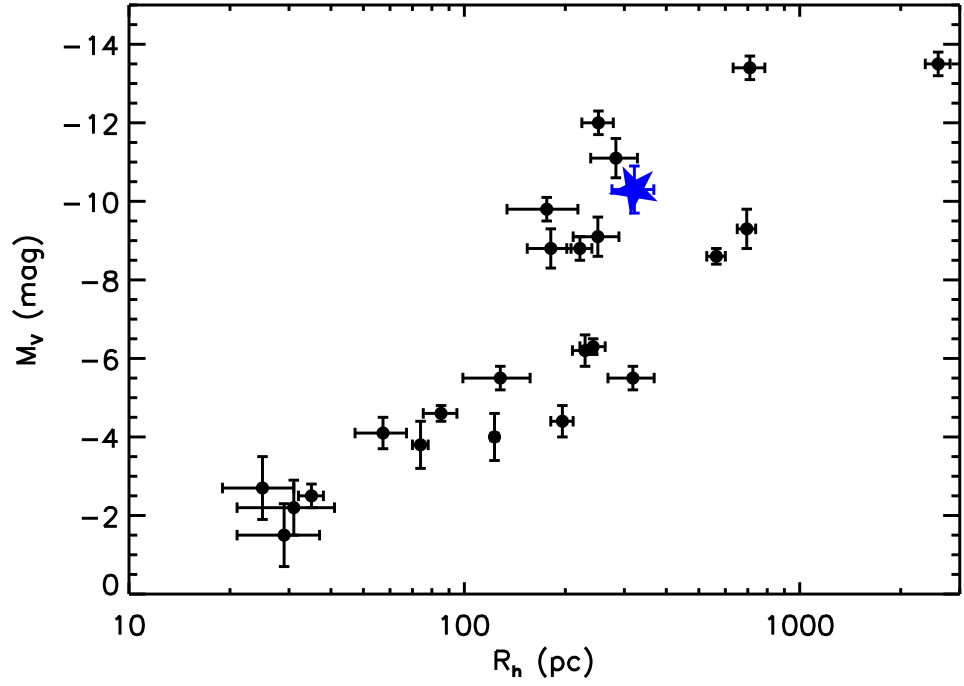}} 
\caption{Absolute magnitude as a function of half light radius for the MW dwarf galaxies (black points with error bars), and Scl-MM-Dw1 (the blue star).  All MW data points were taken from \citet{McConnachie12} or \citet{Sand12}, depending on which had the most up to date numbers.  The properties of Scl-MM-Dw1 are similar to those of the MW dwarfs at its absolute magnitude, particularly Sculptor and Carina.  \label{fig:rhmv}}
\end{center}
\end{figure*}

The size and luminosity of Scl-MM-Dw1 are comparable to those of the MW dwarf galaxies, as can be seen in Figure~\ref{fig:rhmv}.  The MW satellites most similar to Scl-MM-Dw1 are Sculptor ($M_{V}$=$-$11.1 mag; $r_{h}$=283 pc) and Carina ($M_{V}$=$-$9.1 mag; $r_{h}$=250 pc).   Carina in particular also has evidence for young stellar populations \citep[$\sim$0.5-1 Gyr;][]{Monelli03}.

Scl-MM-Dw1 is the first new dwarf galaxy discovered as part of PISCeS, our panoramic imaging campaign to find faint substructure within $r$$\lesssim$150 kpc of both NGC~253 and NGC~5128. Based on initial analyses, our dataset will be able to detect all dwarf galaxies down to $M_{V}$$\sim$$-$8 mag using standard matched-filter techniques \citep[e.g.][]{Koposov08,Walsh09}, allowing for a critical expansion of our knowledge about faint substructure in massive galaxy halos.

\acknowledgments
We thank K.~Hamren for observing help and M.~Conroy, J.~Roll, S.~Moran for their tireless efforts.  DJS thanks Aspen Center for Physics (NSF Grant \#1066293) for their hospitality during paper writing. PG acknowledges NSF grant AST-1010039. E.T. is a Fulbright postdoctoral fellow supported by the Fulbright Program and the Spanish Ministry of Education.

\bibliographystyle{apj}

\begin{thebibliography}{}
\expandafter\ifx\csname natexlab\endcsname\relax\def\natexlab#1{#1}\fi

\bibitem[{{Barnes} {et~al.}(2001){Barnes}, {Staveley-Smith}, {de Blok},
  {Oosterloo}, {Stewart}, {Wright}, {Banks}, {Bhathal}, {Boyce}, {Calabretta},
  {Disney}, {Drinkwater}, {Ekers}, {Freeman}, {Gibson}, {Green}, {Haynes}, {te
  Lintel Hekkert}, {Henning}, {Jerjen}, {Juraszek}, {Kesteven}, {Kilborn},
  {Knezek}, {Koribalski}, {Kraan-Korteweg}, {Malin}, {Marquarding}, {Minchin},
  {Mould}, {Price}, {Putman}, {Ryder}, {Sadler}, {Schr{\"o}der}, {Stootman},
  {Webster}, {Wilson}, \& {Ye}}]{Barnes01}
{Barnes}, D.~G., {Staveley-Smith}, L., {de Blok}, W.~J.~G., {et~al.} 2001,
  \mnras, 322, 486

\bibitem[{{Bellazzini}(2008)}]{bellazzini08}
{Bellazzini}, M. 2008, \memsai, 79, 440

\bibitem[{{Boylan-Kolchin} {et~al.}(2011){Boylan-Kolchin}, {Bullock}, \&
  {Kaplinghat}}]{Boylan11}
{Boylan-Kolchin}, M., {Bullock}, J.~S., \& {Kaplinghat}, M. 2011, \mnras, 415,
  L40

\bibitem[{{Boylan-Kolchin} {et~al.}(2012){Boylan-Kolchin}, {Bullock}, \&
  {Kaplinghat}}]{Boylan12}
---. 2012, \mnras, 422, 1203

\bibitem[{{Chiboucas} {et~al.}(2013){Chiboucas}, {Jacobs}, {Tully}, \&
  {Karachentsev}}]{Chiboucas13}
{Chiboucas}, K., {Jacobs}, B.~A., {Tully}, R.~B., \& {Karachentsev}, I.~D.
  2013, \aj, 146, 126

\bibitem[{{Chiboucas} {et~al.}(2009){Chiboucas}, {Karachentsev}, \&
  {Tully}}]{Chiboucas09}
{Chiboucas}, K., {Karachentsev}, I.~D., \& {Tully}, R.~B. 2009, \aj, 137, 3009

\bibitem[{{Cote} {et~al.}(1997){Cote}, {Freeman}, {Carignan}, \&
  {Quinn}}]{Cote97}
{Cote}, S., {Freeman}, K.~C., {Carignan}, C., \& {Quinn}, P.~J. 1997, \aj, 114,
  1313

\bibitem[{{Dalcanton} {et~al.}(2009){Dalcanton}, {Williams}, {Seth}, {Dolphin},
  {Holtzman}, {Rosema}, {Skillman}, {Cole}, {Girardi}, {Gogarten},
  {Karachentsev}, {Olsen}, {Weisz}, {Christensen}, {Freeman}, {Gilbert},
  {Gallart}, {Harris}, {Hodge}, {de Jong}, {Karachentseva}, {Mateo}, {Stetson},
  {Tavarez}, {Zaritsky}, {Governato}, \& {Quinn}}]{Dalcanton09}
{Dalcanton}, J.~J., {Williams}, B.~F., {Seth}, A.~C., {et~al.} 2009, \apjs,
  183, 67

\bibitem[{{Dotter} {et~al.}(2008){Dotter}, {Chaboyer}, {Jevremovi{\'c}},
  {Kostov}, {Baron}, \& {Ferguson}}]{Dotter08}
{Dotter}, A., {Chaboyer}, B., {Jevremovi{\'c}}, D., {et~al.} 2008, \apjs, 178,
  89

\bibitem[{{Efstathiou} {et~al.}(1992){Efstathiou}, {Bond}, \&
  {White}}]{Efstathiou92}
{Efstathiou}, G., {Bond}, J.~R., \& {White}, S.~D.~M. 1992, \mnras, 258, 1P

\bibitem[{{Girardi} {et~al.}(2010){Girardi}, {Williams}, {Gilbert},
  {Rosenfield}, {Dalcanton}, {Marigo}, {Boyer}, {Dolphin}, {Weisz},
  {Melbourne}, {Olsen}, {Seth}, \& {Skillman}}]{Girardi10}
{Girardi}, L., {Williams}, B.~F., {Gilbert}, K.~M., {et~al.} 2010, \apj, 724,
  1030

\bibitem[{{Grcevich} \& {Putman}(2009)}]{Grcevich09}
{Grcevich}, J., \& {Putman}, M.~E. 2009, \apj, 696, 385

\bibitem[{{Grebel} {et~al.}(2003){Grebel}, {Gallagher}, \&
  {Harbeck}}]{Grebel03}
{Grebel}, E.~K., {Gallagher}, III, J.~S., \& {Harbeck}, D. 2003, \aj, 125, 1926

\bibitem[{{Jaffe} {et~al.}(2001){Jaffe}, {Ade}, {Balbi}, {Bock}, {Bond},
  {Borrill}, {Boscaleri}, {Coble}, {Crill}, {de Bernardis}, {Farese},
  {Ferreira}, {Ganga}, {Giacometti}, {Hanany}, {Hivon}, {Hristov},
  {Iacoangeli}, {Lange}, {Lee}, {Martinis}, {Masi}, {Mauskopf}, {Melchiorri},
  {Montroy}, {Netterfield}, {Oh}, {Pascale}, {Piacentini}, {Pogosyan},
  {Prunet}, {Rabii}, {Rao}, {Richards}, {Romeo}, {Ruhl}, {Scaramuzzi},
  {Sforna}, {Smoot}, {Stompor}, {Winant}, \& {Wu}}]{Jaffe01}
{Jaffe}, A.~H., {Ade}, P.~A., {Balbi}, A., {et~al.} 2001, Physical Review
  Letters, 86, 3475

\bibitem[{{Jerjen} {et~al.}(2000){Jerjen}, {Binggeli}, \& {Freeman}}]{Jerjen00}
{Jerjen}, H., {Binggeli}, B., \& {Freeman}, K.~C. 2000, \aj, 119, 593

\bibitem[{{Jordi} {et~al.}(2006){Jordi}, {Grebel}, \& {Ammon}}]{Jordi06}
{Jordi}, K., {Grebel}, E.~K., \& {Ammon}, K. 2006, \aap, 460, 339

\bibitem[{{Karachentsev} {et~al.}(2004){Karachentsev}, {Karachentseva},
  {Huchtmeier}, \& {Makarov}}]{K04}
{Karachentsev}, I.~D., {Karachentseva}, V.~E., {Huchtmeier}, W.~K., \&
  {Makarov}, D.~I. 2004, \aj, 127, 2031

\bibitem[{{Karachentsev} {et~al.}(2002){Karachentsev}, {Sharina}, {Dolphin},
  {Grebel}, {Geisler}, {Guhathakurta}, {Hodge}, {Karachentseva}, {Sarajedini},
  \& {Seitzer}}]{K02_cen}
{Karachentsev}, I.~D., {Sharina}, M.~E., {Dolphin}, A.~E., {et~al.} 2002, \aap,
  385, 21

\bibitem[{{Klypin} {et~al.}(1999){Klypin}, {Kravtsov}, {Valenzuela}, \&
  {Prada}}]{Klypin99}
{Klypin}, A., {Kravtsov}, A.~V., {Valenzuela}, O., \& {Prada}, F. 1999, \apj,
  522, 82

\bibitem[{{Koposov} {et~al.}(2008){Koposov}, {Belokurov}, {Evans}, {Hewett},
  {Irwin}, {Gilmore}, {Zucker}, {Rix}, {Fellhauer}, {Bell}, \&
  {Glushkova}}]{Koposov08}
{Koposov}, S., {Belokurov}, V., {Evans}, N.~W., {et~al.} 2008, \apj, 686, 279

\bibitem[{{Lee} {et~al.}(1993){Lee}, {Freedman}, \& {Madore}}]{lee93}
{Lee}, M.~G., {Freedman}, W.~L., \& {Madore}, B.~F. 1993, \apj, 417, 553

\bibitem[{{Martin} {et~al.}(2008){Martin}, {de Jong}, \& {Rix}}]{sdssstruct}
{Martin}, N.~F., {de Jong}, J.~T.~A., \& {Rix}, H.-W. 2008, \apj, 684, 1075

\bibitem[{{McConnachie}(2012)}]{McConnachie12}
{McConnachie}, A.~W. 2012, \aj, 144, 4

\bibitem[{{McConnachie} {et~al.}(2009){McConnachie}, {Irwin}, {Ibata},
  {Dubinski}, {Widrow}, {Martin}, {C{\^o}t{\'e}}, {Dotter}, {Navarro},
  {Ferguson}, {Puzia}, {Lewis}, {Babul}, {Barmby}, {Bienaym{\'e}}, {Chapman},
  {Cockcroft}, {Collins}, {Fardal}, {Harris}, {Huxor}, {Mackey},
  {Pe{\~n}arrubia}, {Rich}, {Richer}, {Siebert}, {Tanvir}, {Valls-Gabaud}, \&
  {Venn}}]{McConnachie09}
{McConnachie}, A.~W., {Irwin}, M.~J., {Ibata}, R.~A., {et~al.} 2009, \nat, 461,
  66

\bibitem[{{McLeod} {et~al.}(2006){McLeod}, {Geary}, {Ordway}, {Amato},
  {Conroy}, \& {Gauron}}]{McLeod06}
{McLeod}, B., {Geary}, J., {Ordway}, M., {et~al.} 2006, in Scientific Detectors
  for Astronomy 2005, ed. J.~E. {Beletic}, J.~W. {Beletic}, \& P.~{Amico}, 337

\bibitem[{{Merritt} {et~al.}(2014){Merritt}, {van Dokkum}, \&
  {Abraham}}]{Merritt14}
{Merritt}, A., {van Dokkum}, P., \& {Abraham}, R. 2014, \apjl, 787, L37

\bibitem[{{Monelli} {et~al.}(2003){Monelli}, {Pulone}, {Corsi}, {Castellani},
  {Bono}, {Walker}, {Brocato}, {Buonanno}, {Caputo}, {Castellani}, {Dall'Ora},
  {Marconi}, {Nonino}, {Ripepi}, \& {Smith}}]{Monelli03}
{Monelli}, M., {Pulone}, L., {Corsi}, C.~E., {et~al.} 2003, \aj, 126, 218

\bibitem[{{Moore} {et~al.}(1999){Moore}, {Ghigna}, {Governato}, {Lake},
  {Quinn}, {Stadel}, \& {Tozzi}}]{M99b}
{Moore}, B., {Ghigna}, S., {Governato}, F., {et~al.} 1999, \apjl, 524, L19

\bibitem[{{Percival} {et~al.}(2001){Percival}, {Baugh}, {Bland-Hawthorn},
  {Bridges}, {Cannon}, {Cole}, {Colless}, {Collins}, {Couch}, {Dalton}, {De
  Propris}, {Driver}, {Efstathiou}, {Ellis}, {Frenk}, {Glazebrook}, {Jackson},
  {Lahav}, {Lewis}, {Lumsden}, {Maddox}, {Moody}, {Norberg}, {Peacock},
  {Peterson}, {Sutherland}, \& {Taylor}}]{Percival01}
{Percival}, W.~J., {Baugh}, C.~M., {Bland-Hawthorn}, J., {et~al.} 2001, \mnras,
  327, 1297

\bibitem[{{Radburn-Smith} {et~al.}(2011){Radburn-Smith}, {de Jong}, {Seth},
  {Bailin}, {Bell}, {Brown}, {Bullock}, {Courteau}, {Dalcanton}, {Ferguson},
  {Goudfrooij}, {Holfeltz}, {Holwerda}, {Purcell}, {Sick}, {Streich}, {Vlajic},
  \& {Zucker}}]{ghosts}
{Radburn-Smith}, D.~J., {de Jong}, R.~S., {Seth}, A.~C., {et~al.} 2011, \apjs,
  195, 18

\bibitem[{{Rizzi} {et~al.}(2007){Rizzi}, {Tully}, {Makarov}, {Makarova},
  {Dolphin}, {Sakai}, \& {Shaya}}]{rizzi07}
{Rizzi}, L., {Tully}, R.~B., {Makarov}, D., {et~al.} 2007, \apj, 661, 815

\bibitem[{{Sand} {et~al.}(2012){Sand}, {Strader}, {Willman}, {Zaritsky},
  {McLeod}, {Caldwell}, {Seth}, \& {Olszewski}}]{Sand12}
{Sand}, D.~J., {Strader}, J., {Willman}, B., {et~al.} 2012, \apj, 756, 79

\bibitem[{{Schlafly} \& {Finkbeiner}(2011)}]{Schlafly11}
{Schlafly}, E.~F., \& {Finkbeiner}, D.~P. 2011, \apj, 737, 103

\bibitem[{{Spergel} {et~al.}(2007){Spergel}, {Bean}, {Dor{\'e}}, {Nolta},
  {Bennett}, {Dunkley}, {Hinshaw}, {Jarosik}, {Komatsu}, {Page}, {Peiris},
  {Verde}, {Halpern}, {Hill}, {Kogut}, {Limon}, {Meyer}, {Odegard}, {Tucker},
  {Weiland}, {Wollack}, \& {Wright}}]{Spergel07}
{Spergel}, D.~N., {Bean}, R., {Dor{\'e}}, O., {et~al.} 2007, \apjs, 170, 377

\bibitem[{{Springel} {et~al.}(2006){Springel}, {Frenk}, \&
  {White}}]{Springel06}
{Springel}, V., {Frenk}, C.~S., \& {White}, S.~D.~M. 2006, \nat, 440, 1137

\bibitem[{{Stetson}(1994)}]{Stetson94}
{Stetson}, P.~B. 1994, \pasp, 106, 250

\bibitem[{{Walsh} {et~al.}(2009){Walsh}, {Willman}, \& {Jerjen}}]{Walsh09}
{Walsh}, S.~M., {Willman}, B., \& {Jerjen}, H. 2009, \aj, 137, 450

\end{thebibliography}

\clearpage

\clearpage

\begin{deluxetable*}{lcccccccccc}
\tablecolumns{2}
\tablecaption{Properties of Scl-MM-Dw1 \label{table:properties}}
\tablehead{
\colhead{Parameter}  & \colhead{Value} \\
}\\
\startdata
$m-M$ (mag) & 27.98$\pm$0.32\\
D (Mpc) &  3.9$\pm$0.5\\
$M_{V}$ (mag) & $-$10.3$\pm$0.6 \\
RA (h:m:s) & 00:47:34.93 $\pm$2"\\
DEC (d:m:s) & -26:23:19.7 $\pm$2"\\
$r_{h}$ (arcsec) & 16.8$\pm$2.4 \\
$r_{h}$ (pc) & 340$\pm$50 \\
$\epsilon$ & $<$0.42\tablenotemark{a}\\
\enddata
\tablenotetext{a}{$\epsilon$ corresponds to the 95\% upper confidence limit.}

\end{deluxetable*}


\end{document}